\documentclass[journal=nalefd,manuscript=article,layout=traditional]{achemso}

\usepackage[version=3]{mhchem} 
\usepackage{natbib}
\usepackage{color}

\author{P.\,Sessi} 
\email{paolo.sessi@physik.uni-wuerzburg.de}
\affiliation{Physikalisches Institut, Experimentelle Physik 2, Universit{\"a}t W{\"u}rzburg, 
Am Hubland, 97074 W{\"u}rzburg, Germany}
\author{T.\,Bathon}
\affiliation{Physikalisches Institut, Experimentelle Physik 2, Universit{\"a}t W{\"u}rzburg, 
Am Hubland, 97074 W{\"u}rzburg, Germany}
\author{K.\,A.\,Kokh} 
\affiliation{V.S.\,Sobolev Institute of Geology and Mineralogy, Siberian Branch, 
Russian Academy of Sciences,630090 Novosibirsk,  Russia}
\alsoaffiliation{Novosibirsk State University, 630090 Novosibirsk, Russia}
\affiliation{Novosibirsk State University, 630090 Novosibirsk, Russia}
\author{O.\,E.\,Tereshchenko} 
\affiliation{A.V. Rzanov Institute of Semiconductor Physics, Siberian Branch, 
Russian Academy of Sciences, 630090 Novosibirsk,  Russia}
\alsoaffiliation{Novosibirsk State University, 630090 Novosibirsk, Russia}
\author{M.\,Bode}
\affiliation{Physikalisches Institut, Experimentelle Physik 2, Universit{\"a}t W{\"u}rzburg, 
Am Hubland, 97074 W{\"u}rzburg, Germany}
\alsoaffiliation{Wilhelm Conrad R{\"o}ntgen-Center for Complex Material Systems (RCCM), 
Universit{\"a}t W{\"u}rzburg, Am Hubland, 97074 W{\"u}rzburg, Germany}

\title{Probing the electronic properties of individual MnPc molecules coupled to topological states}


\begin{document}


\newpage

\begin{abstract}
Hybrid organic/inorganic interfaces have been widely reported 
to host emergent properties that go beyond those of their single constituents. 
Coupling molecules to the recently discovered topological insulators, 
which possess a linearly dispersing and spin-momentum--locked Dirac fermions,
may offer a promising platform towards new functionalities.  
Here, we report a scanning tunneling microscopy and spectroscopy study 
of the prototypical interface between MnPc molecules and a Bi$_2$Te$_3$ surface. 
MnPc is found to bind stably to the substrate through its central Mn atom. 
The adsorption process is only accompanied with a minor charge transfer across the interface, 
resulting in a moderately n-doped Bi$_2$Te$_3$ surface. 
More remarkably, topological states remain completely unaffected by the presence of the molecules, 
as evidenced by the absence of scattering patterns around adsorption sites. 
Interestingly, we show that, while the HOMO and LUMO orbitals closely resembles 
those of MnPc in the gas phase, a new hybrid states emerges through interaction with the substrate. 
Our results pave the way towards hybrid organic--topological insulator heterostructures, 
which may unveil a broad range of exciting and unknown phenomena. 
\end{abstract}
\newpage

\section{Introduction}
Over the past decade the emerging field of molecular electronics has shown its great potential 
for the realization of sensors, switches, and magnetic memories \cite{H2009}. 
In comparison with other conventional semiconductor-based integrated circuits, 
molecular electronics holds the promise of several advantages, 
such as low production costs, compatibility with scaling, low power consumption and, 
most of all, high tunability through chemical modifications. 
Although the choice of appropriate substrates and the formation of reproducible electrical leads
to support and contact single molecules have been the main challenges 
towards a broader application of molecular electronics, respectively, 
the identification of suitable materials may open up effective strategies 
for further expanding the functionality of molecular building blocks.

Within this framework, molecules have been successfully coupled to a large variety of materials 
to create efficient spin valves \cite{XVV2004,RGB2005,CHV2009}, as well as Coulomb blockade \cite{PPG2002}, 
negative differential resistance effects \cite{CRR1999} and molecular switches \cite{ASF2012}. 
On this route, the recently discovered topological insulators (TI) 
represent a promising new class of materials \cite{HQW2008,CAC2009}. 
TIs  are insulating in the bulk but conductive on their surface,  
where they host linearly dispersing Dirac fermions protected by time-reversal symmetry. 
The strong spin-orbit coupling perpendicularly locks the spin to the momentum, 
resulting in a chiral spin texture which forbids backscattering, thus increasing the spin coherence time 
and making charge currents intrinsically related to spin currents \cite{HK2010}.  
All these properties make TIs a particularly attractive substrate for molecular magnetism.

Despite these interesting properties, we are not aware 
of any investigation that discusses the structural or electronic properties 
of interfaces between molecules and topo\-logical insulators. 
Here, we report on a detailed scanning tunneling microscopy (STM) and spectroscopy (STS) study 
of a prototypical model system consisting of Mn-phthalocyanine (MnPc) coupled to Bi$_2$Te$_3$.
Our choice was motived by the fact that transition metal phthalocyanines are representative 
of a broad class of planar metal-organic compounds where the presence of partially empty $d$ orbitals 
in the central metal atom gives them interesting magnetic properties \cite{FSP2011,SSP2012,SWF2014}.  
Bi$_2$Te$_3$, on the other hand,  is a well studied TI characterized by a single Dirac cone 
centered around the $\Gamma$ point which makes it a model system 
for the large class of chalcogenide TIs \cite{CAC2009,SOB2013}.

By analyzing both the morphology as well as the electronic properties 
of the interface down to the single molecule level, we provide evidence 
that MnPc stably binds through its central transition metal atom to Bi$_2$Te$_3$. 
This process takes place with a minor interfacial charge transfer.
Correspondingly, the topological states are largely unaffected by the presence of the molecules, 
as evidenced by the moderate band bending taking place at the interface and, 
more remarkably, by the absence of scattering patterns around MnPc molecules.
The highest occupied molecular orbital (HOMO) and the lowest unoccupied molecular orbital (LUMO) 
closely resemble those of the gas phase, as observed for molecules adsorbed on insulators \cite{USR2013}, 
thereby confirming their weak interaction with the substrate. 
Nevertheless, similar to Pc molecules coupled to metallic surfaces \cite{ZFS2010}, 
a new hybrid molecule--Bi$_2$Te$_3$ state emerges close to the Fermi level.

\section{Results and discussion}

\begin{figure}[t]   
\begin{minipage}[t]{0.65\textwidth} 
\includegraphics[width=10cm]{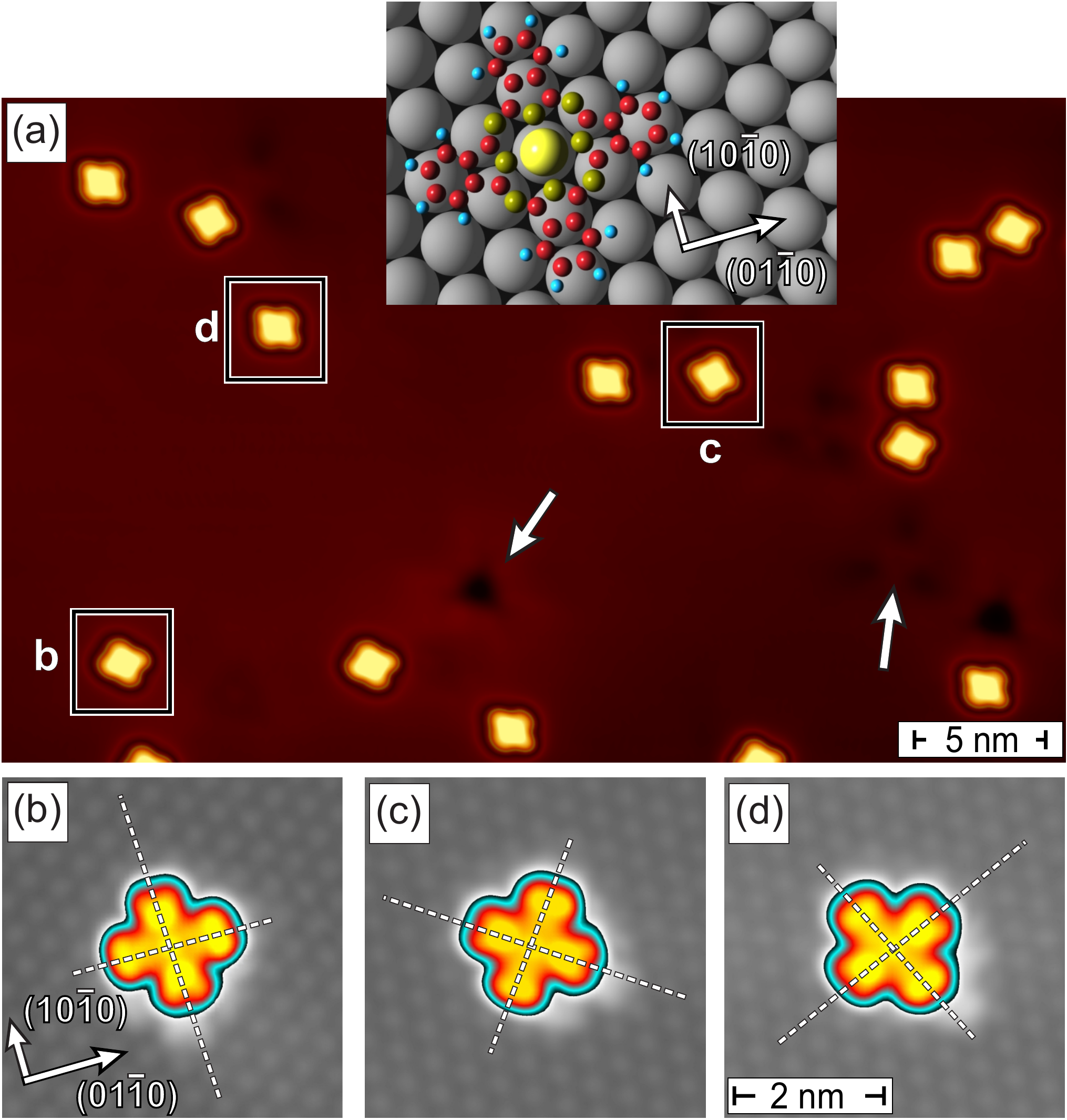}%
\end{minipage}
\hfill
\begin{minipage}[b]{0.34\textwidth}
\caption{ (a) Constant-current image of MnPc molecules on Bi$_2$Te$_3$. 
	Typical surface defects of the TI are indicated by arrows. 
	Three different molecule orientation are visible. 
	(b-d) Atomically resolved images showing that the central Mn atom of the molecule 
	is always located on top of a Te atom, irrespective of the molecular orientation. 
	The symmetry of the substrate results in three different molecule orientations 
	and in a reduction of the Mn symmetry from 4-fold to 2-fold (see inset).  Scanning parameters I = 20 pA, V = -0.2 V.}
\label{fgr:AdsSite}
\end{minipage}
\end{figure}   
Since the interfacial properties strongly depend on the interaction of the molecules with the underlying substrate, 
a detailed knowledge of the adsorption geometry and bond configuration is mandatory. 
\ref{fgr:AdsSite}(a) shows a topographic image of MnPc molecules 
thermally grown on Bi$_2$Te$_3$ with the substrate kept at 5.3\,K. 
Interpolation of the Te lattice observed in the atomically resolved images reported in \ref{fgr:AdsSite}(b-d) 
onto the molecule center reveals that, independent of the molecular orientation, 
the central Mn atom always sits on top of an underlying Te atom. 

If the bonding process would be strongly dominated by the central Mn ion, 
any orientation between the molecular axes and the substrate would be possible. 
However, only three equivalent orientations of the molecules are visible in \ref{fgr:AdsSite}(a)-(d). 
While two isoindole groups are oriented in $\langle 10{\bar 1}0 \rangle$ directions of the substrate 
i.e.\ along dense-packed Te rows of the Bi$_2$Te$_3$ surface,
the other two groups are oriented in $\langle 01{\bar 1}0 \rangle$ directions.  
This observation provides evidence that the ligand interacts with the substrate thus favoring a specific alignment. 
The development of three eqivalent molecular orientation 
is the result of the combination of the four-fold symmetry of the MnPc 
with the six-fold symmetric (0001) surface of the Bi$_2$Te$_3$ substrate. 
As schematically illustrated for the undistorted molecule in the inset of \ref{fgr:AdsSite}(a),
the adsorption geometry extracted from \ref{fgr:AdsSite}(b-d) 
results in two inequivalent adsorption environments for the isoindole groups,
which significantly impact the molecule appearance and break its four-fold symmetry. 

In fact, a slight distortion of the molecule's four-fold symmetry 
can already be recognized in the overview scan of \ref{fgr:AdsSite}(a).  
This is confirmed by the atomically resolved images 
reported in \ref{fgr:AdsSite}(b-d) which also show the inequivalent appearance of the axis, 
with the one aligned along a dense-packed rows of Te atoms appearing slightly brighter than the other. 
This indicates that the symmetry of the molecules 
is reduced to two-fold upon adsorption \cite{CKB2008,CCW2010,BL2013}.  
As it will be described in the following, the ligand--substrate interaction impacts 
the electronic states and leads to the appearance of hybrid molecule--substrate states. 

Interestingly, this adsorption scenario was found to be temperature independent up to room temperature.  
By warming the sample up to room temperature and subsequently cooling it down at the measurement temperature (see methods)
no tendency towards surface diffusion and clustering could be detected (data not shown here). 
This observation points towards a relatively strong bond between the molecule and the substrate.  
Similar findings have been widely reported for Pc molecules 
adsorbed on different substrates \cite{PSA2004,CGD2007,MLO2010,MKR2011,MRK2012}. 
For MnPc, this has been attributed to the presence of empty $d$-orbitals 
localized on the central atom which point perpendicular to the surface plane. 
In the present case, they can thus effectively hybridize with the substrate $p_z$ orbitals, 
which are known to be those closer to the Fermi level \cite{ZLQ2009}.

\begin{figure}[t]   
\includegraphics[width=\columnwidth]{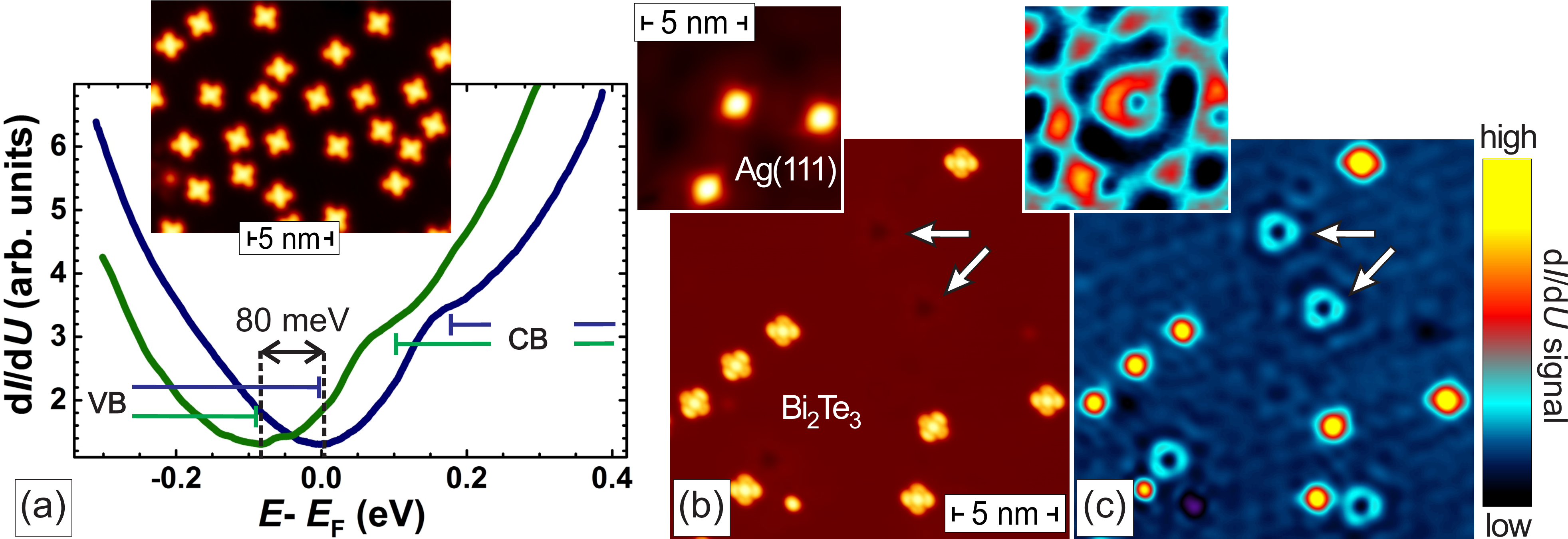}%
\caption{(a) STS spectra taken on pristine Bi$_2$Te$_3$ (blue line) and after MnPc deposition (green line). 
	The coverage corresponds to approximately 1/4 of a monolayer (see inset). 
	VB and CB refer to valence and conduction band, respectively. 
	Constant-current and simultaneously acquired differential conductance map 
	for MnPc on Bi$_2$Te$_3$ (b,c) and Ag(111) (insets). 
	While MnPc strongly scatters $s-p$ derived surface states of Ag(111), 
	it leaves topological states unaffected. This is in contrast to Bi$_2$Te$_3$ pristine surface defects, which strongly scatters the topological state as evidenced by the standing wave pattern surrounding them (see arrows). Scanning parameters I = 20 pA, (a) V = -0.3 V (b,c) V = +0.6 V.   }
\label{fgr:Spec}
\end{figure}   
Earlier studies of molecules strongly bound to metal \cite{MLO2010,MKR2011} 
or semiconductor surfaces \cite{PSA2004} provided compelling evidence 
that stable molecule--substrate bonds 
may strongly influence the single molecule electronic properties. 
In oder to investigate whether the creation of the MnPc/Bi$_2$Te$_3$ interface 
gives rise to any significant charge redistribution 
we have investigated the local density of states as it can be inferred by STS measurements.
\ref{fgr:Spec}(a) shows STS spectra taken on Bi$_2$Te$_3$ 
before (green line) and after (blue line) molecules deposition.
The MnPc coverage approximately amounts to 1/4 of a monolayer (see STM image in the inset). 
The energy levels positioning was done according to the procedure described in Ref.\citenum{SOB2013}.
Comparison shows that MnPc deposition leads to a 80\,meV rigid shift of the spectra towards negative energies,
indicative of a very moderate n-doping of the Bi$_2$Te$_3$ surface. 
This value is well below those reported for 3$d$ single transition metal adatoms on TIs, 
where energy shifts of more than 100\,meV were observed 
already for coverages of only few percent of a monolayer \cite{SBK2013}.  

Even more remarkably, differential conductance $\mathrm{d}I/\mathrm{d}U$ maps indicate 
that MnPc has little to no influence on the propagation of topological surface states.  
This is evidenced by the data of \ref{fgr:Spec}(b) and (c) which report 
a constant-current image and the simultaneously acquired $\mathrm{d}I/\mathrm{d}U$ map 
obtained at an energy of $+600$\,meV above the Fermi level, respectively. 
Contrary to surface defects usually found on the Bi$_2$Te$_3$ surface (see arrows) 
which strongly scatter the surface states as signaled by the presence a standing wave pattern 
surrounding them [cf.\ \ref{fgr:Spec}(c)], no standing waves are visible around the MnPc molecules. 
The absence of electron interference patterns around MnPc molecules is a direct consequence of the absence of any significant scattering potential  
associated to their presence of the surface Bi$_2$Te$_3$.

In order to emphasize this quite unusual situation, the insets of \ref{fgr:Spec}(b) and (c) 
show the topography and a $\mathrm{d}I/\mathrm{d}U$ map 
obtained for the same molecules adsorbed on Ag(111), respectively. 
In this case the topologically trivial $s$-$p$--derived surface state of Ag(111) strongly scatters at MnPc molecules,
resulting in a pronounced oscillatory pattern of the local density of states. 
This also implies that MnPc might serve as an effective passivation layer 
without affecting the transport properties of the topological states.

\begin{figure*}[t]   
\includegraphics[width=0.9\columnwidth]{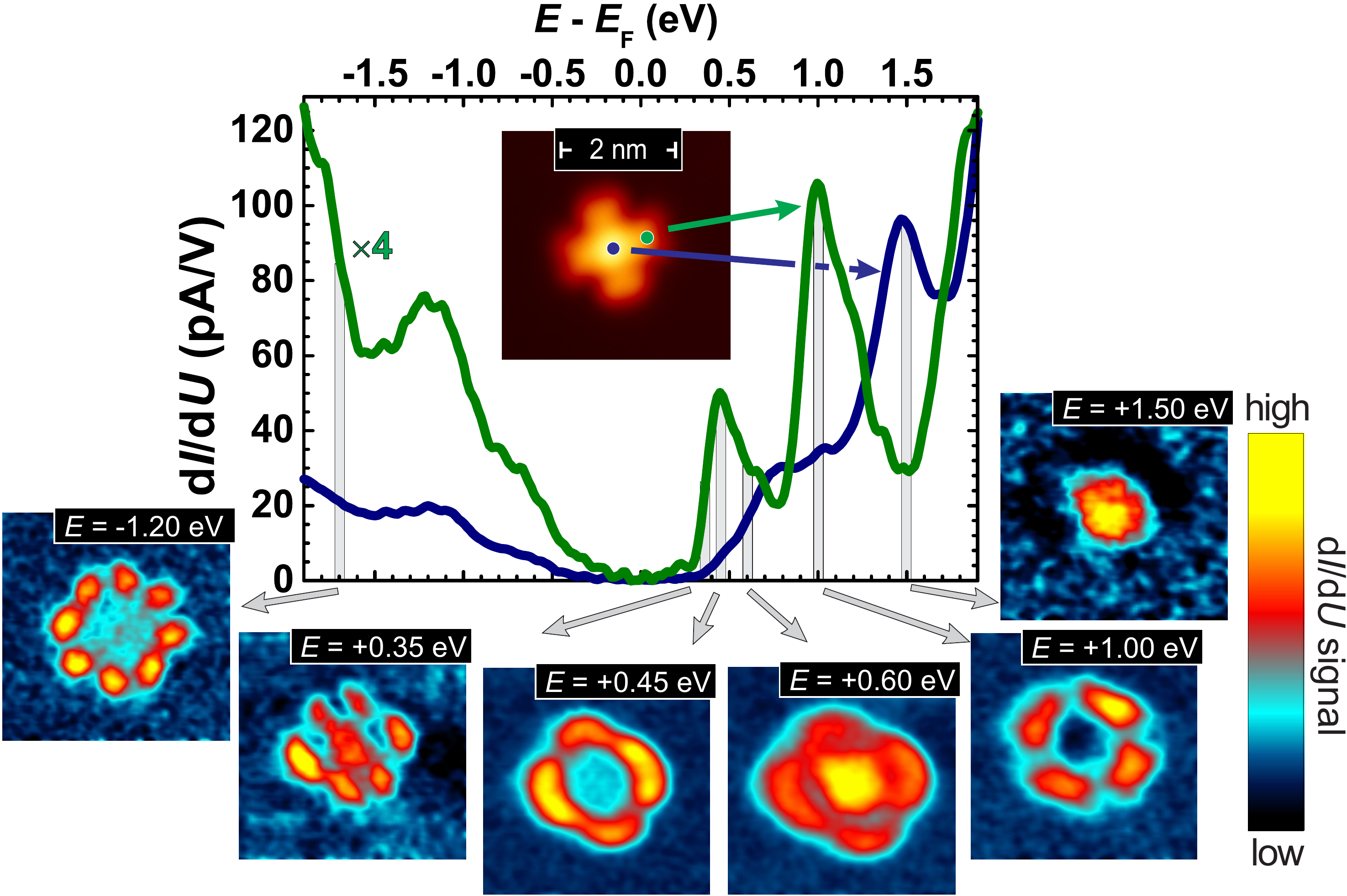}%
\caption{STS spectra taken on top of a ligand (green line) and on the Mn atom (blue). 
	Insets report differential conductance maps acquired in constant-current mode at significant energies, 
	which display the spatial distribution of the states. STS stabilization parameters I = 20 pA, V = -2V. The same set-point current has been used to acquired the dI/dU maps displayed in the insets. }
\label{fgr:Ligand}
\end{figure*}   
To shed light on their interaction with the substrate we also analyzed 
the energy and spatial distribution of the molecular orbitals. 
\ref{fgr:Ligand} reports $\mathrm{d}I/\mathrm{d}U$ curves obtained 
by positioning the tip over a ligand (green line) or over the central Mn atom (blue).   
All energies are referred with respect to the Fermi level. 
Three strong peaks are visible over the ligand, one in the occupied states at $-1.20$\,eV, 
the other two in the unoccupied states at $+0.45$ and $+1.00$\,eV, respectively.  
Differential conductance $\mathrm{d}I/\mathrm{d}U$ maps taken at energies corresponding to these peaks 
allow to visualize the intramolecular spatial distribution of the responsible electronic states (see insets).  
The states at $-1.20$\,eV and $+1.00$\,eV closely resemble the HOMO ($a_{1u}$ symmetry) 
and LUMO ($e_g$ symmetry) orbital density contours of Pc molecules in the gas phase, respectively. 
Similar observations have been reported for molecules adsorbed on insulators \cite{USR2013} 
and indicate that these states are weakly interacting with the substrate. 

The conductance at $+0.45$\,eV, on the other hand, exhibits a two fold symmetry, 
as already observed in the constant-current image of \ref{fgr:AdsSite}. 
This allows us to assign this feature to a new molecule--surface hybrid state 
which may occur at the interface where the molecule interacts with metallic states \cite{ZFS2010}.
Our assignment is corroborated by the spatial distribution of this state, 
which very closely resembles hybrid states found at Pc/noble metal interfaces \cite{LLY2010}. 
It is further confirmed by the differential conductance map taken 
at the onset of this resonance (see panel $E = -0.35$\,eV), 
which shows how this state emerges from a 6-fold symmetry, i.e.\ the symmetry of the substrate. 

The spectrum measured with the STM tip located above the central Mn ion shows 
that all peaks discussed so far are strongly suppressed (see blue spectrum). 
Instead, a step-like function is visible in the energy region within the hybrid state 
and the LUMO (i.e.\ between 0.5 and 1\,eV) which indicates 
the presence of several orbitals in a narrow energy region.
Indeed, inspection of the differential conductance map measured at $E = +0.60$\,eV 
confirms the presence of an electronic state that is mainly localized at the Mn atom. 
It is well known that---contrary to ``late'' transition metals (Ni and Cu) Pc molecules, 
where the 3$d$ orbital of the central atom does not play any significant role 
in determining the low energy electronic properties---the situation is quite different 
for ``early'' transition metal Pc (Fe and Mn) \cite{MKR2011}.  
In particular, for MnPc both the HOMO and the LUMO as well as 3$d$-orbitals 
with symmetry $e_{g} (d_{xz,yz})$, $a_{1g} (d_{z^2})$, and $b_{2g} (d_{xy})$ 
can be found within a relatively narrow energy range around the Fermi level \cite{GML2011}. 
This results in a stronger metal--ligand hybridization \cite{MRK2012}, 
with the delocalized orbitals acquiring a $d$ component, 
an effect which depends on the spatial and energy proximity of the states with  $e_g$ symmetry. 

At an energy of approximately 1.5\,eV, a strong peak absent on the ligand 
emerges from the background once the tip is positioned onto the Mn atom. 
Although a detailed characterization of the Mn orbitals occupancy 
is beyond the scope of the present work, 
its spatial location and strong intensity allows to assign it to the $d_{z^2}$ orbital, whose density, 
perpendicular to surface plane, extends into the vacuum 
making its detection by the STM tip particularly effective. 
Similar to MnPc molecules adsorbed on Bi(110), 
its broadening is though to be a result of the hybridization with the substrate \cite{SSP2012}.

\begin{figure}[t]   
\includegraphics[width=\columnwidth]{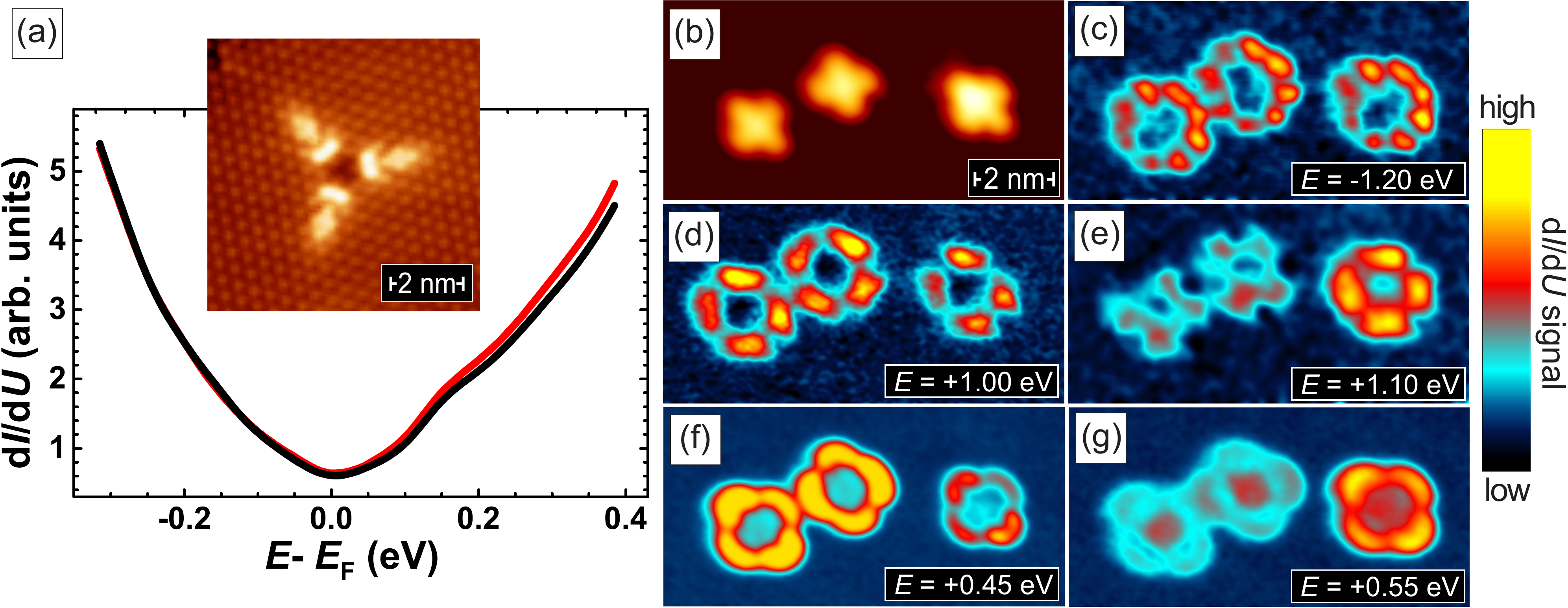}%
\caption{(a) STS spectra taken on a typical Bi$_2$Te$_3$ defect (red line) 
	as displayed in the inset compared with a STS taken on the unperturbed surface (black line).  
	(b) constant-current image of three MnPc molecules. 
	The left one has  the central Mn atom sitting on top of the defect arm. 
	(c-g) differential conductance maps for HOMO, LUMO and hybrid state. Stabilization parameters I = 20 pA, V = -0.3V. The same set-point current has been used to acquired the dI/dU maps displayed in the insets.}
\label{fgr:Defect}
\end{figure}   
Interestingly, we found clear signature of how electronic correlations on the central 3d metal ion influence the overall electronic structure of the system.
This is done by studying the MnPc energy levels once the central Mn atom is adsorbed on top of a defect, which may be view as a perturbation potential influencing the e-e- interaction.
Fig. \ref{fgr:Defect}(a) reports a constant-current image (inset) and STS spectra 
of the pristine Bi$_2$Te$_3$ surface and a typical Bi$_2$Te$_3$ defect, 
i.e.\ a Bi$_{\rm Te}$ antisite located in the fifth layer \cite{WZS2011}.
Note that the influence of the defect on the surface electronic properties is marginal, 
as can be inferred by comparing spectra taken on top of the defect (red line) 
and a defect-free region few nanometers away from it (black line).
Whenever a molecule was found adsorbed with the central Mn atom sitting exactly on top of a defect, the position of its energy levels resulted strongly influenced. Nevertheless,  energy levels stay unchanged if the ligand happens to be located over the defect. 
For example, fig.\ref{fgr:Defect}(b) shows a constant-current image 
displaying three MnPc molecules, in which the right one has the Mn sitting on a defect. 
Differential conductance maps measured at some characteristic energies are reported in panels (c-g).
While no difference can detected for the HOMO [see fig. \ref{fgr:Defect}(c)], thus indicating its pure ligand origin, 
both the maximum intensity of the LUMO as well as the hybrid states 
is shifted by approximately $+100$\,meV for molecules sitting on the defects,
as shown in fig. \ref{fgr:Defect}(d,e) and fig. \ref{fgr:Defect}(f,g), respectively.

Our findings demonstrate that MnPc molecules can be effectively coupled to TIs. 
Although the molecule--substrate bond results to be strong, 
bond formation is only associated with minor charge transfer across the interface. 
Remarkably, the adsorption of MnPc molecules leaves the topological properties 
of the Bi$_2$Te$_3$ surface states unaffected, 
as evidenced by the absence of characteristic interference patterns. 
The analysis of the molecular orbitals allows to visualize HOMO and LUMO orbitals, 
closely resembling those of molecules in the gas phase.
At the same time, a hybrid molecule--substrate state emerges into the unoccuppied states.
Overall, our characterization of a prototype hybrid organic--TI interface provides evidence 
that these materials can be successfully combined to create stable interfaces. 
Because of the non trivial spin-properties of topological states, 
these interfaces represent thus a promising platform to explore spin-related phenomena 
once magnetic-molecules are coupled to them. For example, the long spin relaxation time typical of molecules and the absence of backscattering in topological states  is expected to result in interfaces with a strongly increased spin coherence time, an important quantity for spintronic devices in which the information is encoded by the spin degree of freedom.

\section{Methods}
The experiments have been performed 
in a ultra high vacuum (UHV) system equipped with a cryogenic STM.   
The Bi$_2$Te$_3$ single-crystal samples were grown using the Bridgman technique \cite{KMG2014}.
They have been cleaved in UHV at a base pressure of $3 \cdot 10^{-11}$ mbar 
and immediately inserted into the STM operated at $T = 4.8$\,K. 
Source materials with 99\,\% MnPc (Sigma Aldrich) 
were first purified by sublimation in a tube furnace 
and then degassed overnight under UHV condition.
MnPc molecules were deposited directly onto the cold Bi$_2$Te$_3$ surface ($T = 5.3$\,K)  
by using an home made Knudsen cell.  

\begin{acknowledgement}

This work has been funded within SPP 1666 ``Topologische Isolatoren'' (project BO 1468/21-1).
K.A.K. and O.E.T. acknowledge financial support by the RFBR (Grant Nos.\ 13-02-92105 and 14-08-31110).

\end{acknowledgement}





\begin{mcitethebibliography}{31}
\providecommand*{\natexlab}[1]{#1}
\providecommand*{\mciteSetBstSublistMode}[1]{}
\providecommand*{\mciteSetBstMaxWidthForm}[2]{}
\providecommand*{\mciteBstWouldAddEndPuncttrue}
  {\def\EndOfBibitem{\unskip.}}
\providecommand*{\mciteBstWouldAddEndPunctfalse}
  {\let\EndOfBibitem\relax}
\providecommand*{\mciteSetBstMidEndSepPunct}[3]{}
\providecommand*{\mciteSetBstSublistLabelBeginEnd}[3]{}
\providecommand*{\EndOfBibitem}{}
\mciteSetBstSublistMode{f}
\mciteSetBstMaxWidthForm{subitem}{(\alph{mcitesubitemcount})}
\mciteSetBstSublistLabelBeginEnd{\mcitemaxwidthsubitemform\space}
{\relax}{\relax}

\bibitem[Heat(2009)]{H2009}
Heat,~J.~R. \emph{Annu.\ Rev.\ Mater.\ Res.} \textbf{2009}, \emph{39}, 1\relax
\mciteBstWouldAddEndPuncttrue
\mciteSetBstMidEndSepPunct{\mcitedefaultmidpunct}
{\mcitedefaultendpunct}{\mcitedefaultseppunct}\relax
\EndOfBibitem
\bibitem[Xiong et~al.(2004)Xiong, Wu, Vardeny, and Shi]{XVV2004}
Xiong,~Z.~H.; Wu,~D.; Vardeny,~Z.~V.; Shi,~J. \emph{Nature} \textbf{2004},
  \emph{427}, 821\relax
\mciteBstWouldAddEndPuncttrue
\mciteSetBstMidEndSepPunct{\mcitedefaultmidpunct}
{\mcitedefaultendpunct}{\mcitedefaultseppunct}\relax
\EndOfBibitem
\bibitem[Rocha et~al.(2005)Rocha, Garc{\'i}a-Su{\'a}rez, Bailey, Lambert,
  Ferrer, and Sanvito]{RGB2005}
Rocha,~A.~R.; Garc{\'i}a-Su{\'a}rez,~V.~M.; Bailey,~S.~W.; Lambert,~C.~J.;
  Ferrer,~J.; Sanvito,~S. \emph{Nature Mater.} \textbf{2005}, \emph{4},
  335\relax
\mciteBstWouldAddEndPuncttrue
\mciteSetBstMidEndSepPunct{\mcitedefaultmidpunct}
{\mcitedefaultendpunct}{\mcitedefaultseppunct}\relax
\EndOfBibitem
\bibitem[Cinchetti et~al.(2009)Cinchetti, Heimer, W{\"u}stenberg, Andreyev,
  Bauer, Lach, Ziegler, Gao, and Aeschlimann]{CHV2009}
Cinchetti,~M.; Heimer,~K.; W{\"u}stenberg,~J.-P.; Andreyev,~O.; Bauer,~M.;
  Lach,~S.; Ziegler,~C.; Gao,~Y.; Aeschlimann,~M. \emph{Nature Mater.}
  \textbf{2009}, \emph{8}, 115\relax
\mciteBstWouldAddEndPuncttrue
\mciteSetBstMidEndSepPunct{\mcitedefaultmidpunct}
{\mcitedefaultendpunct}{\mcitedefaultseppunct}\relax
\EndOfBibitem
\bibitem[Park et~al.(2002)Park, Pasupathy, Goldsmith, Chang, Yaish, Petta,
  Rinkoski, Sethna, Abru{\~n}a, McEuen, and Ralph]{PPG2002}
Park,~J.; Pasupathy,~A.~N.; Goldsmith,~J.~I.; Chang,~C.; Yaish,~Y.;
  Petta,~J.~R.; Rinkoski,~M.; Sethna,~J.~P.; Abru{\~n}a,~H.~D.; McEuen,~P.~L.;
  Ralph,~D.~C. \emph{Nature} \textbf{2002}, \emph{417}, 722\relax
\mciteBstWouldAddEndPuncttrue
\mciteSetBstMidEndSepPunct{\mcitedefaultmidpunct}
{\mcitedefaultendpunct}{\mcitedefaultseppunct}\relax
\EndOfBibitem
\bibitem[Chen et~al.(1999)Chen, Reed, Rawlett, and Tour]{CRR1999}
Chen,~J.; Reed,~M.~A.; Rawlett,~A.~M.; Tour,~J.~M. \emph{Science}
  \textbf{1999}, \emph{286}, 5444\relax
\mciteBstWouldAddEndPuncttrue
\mciteSetBstMidEndSepPunct{\mcitedefaultmidpunct}
{\mcitedefaultendpunct}{\mcitedefaultseppunct}\relax
\EndOfBibitem
\bibitem[Auw{\"a}rter et~al.(2012)Auw{\"a}rter, Seufert, Bischoff, Ecija,
  Vijayaraghavan, Joshi, Klappenberger, Samudrala, and Barth]{ASF2012}
Auw{\"a}rter,~W.; Seufert,~K.; Bischoff,~F.; Ecija,~D.; Vijayaraghavan,~S.;
  Joshi,~S.; Klappenberger,~F.; Samudrala,~N.; Barth,~J.~V. \emph{Nature
  Nanotechn.} \textbf{2012}, \emph{7}, 41\relax
\mciteBstWouldAddEndPuncttrue
\mciteSetBstMidEndSepPunct{\mcitedefaultmidpunct}
{\mcitedefaultendpunct}{\mcitedefaultseppunct}\relax
\EndOfBibitem
\bibitem[Hsieh et~al.(2008)Hsieh, Qian, Wray, Xia, Hor, Cava, and
  Hasan]{HQW2008}
Hsieh,~D.; Qian,~D.; Wray,~L.; Xia,~Y.; Hor,~Y.~S.; Cava,~R.~J.; Hasan,~M.~Z.
  \emph{Nature} \textbf{2008}, \emph{452}, 970\relax
\mciteBstWouldAddEndPuncttrue
\mciteSetBstMidEndSepPunct{\mcitedefaultmidpunct}
{\mcitedefaultendpunct}{\mcitedefaultseppunct}\relax
\EndOfBibitem
\bibitem[Chen et~al.(2009)Chen, Analytis, Chu, Liu, Mo, Qi, Zhang, Lu, Dai,
  Fang, Zhang, Fisher, Hussain, and Shen]{CAC2009}
Chen,~Y.~L.; Analytis,~J.~G.; Chu,~J.-H.; Liu,~Z.~K.; Mo,~S.-K.; Qi,~X.~L.;
  Zhang,~H.~J.; Lu,~D.~H.; Dai,~X.; Fang,~Z.; Zhang,~S.~C.; Fisher,~I.~R.;
  Hussain,~Z.; Shen,~Z.-X. \emph{Science} \textbf{2009}, \emph{325}, 178\relax
\mciteBstWouldAddEndPuncttrue
\mciteSetBstMidEndSepPunct{\mcitedefaultmidpunct}
{\mcitedefaultendpunct}{\mcitedefaultseppunct}\relax
\EndOfBibitem
\bibitem[Hasan and Kane(2010)]{HK2010}
Hasan,~M.~Z.; Kane,~C.~L. \emph{Rev. Mod. Phys.} \textbf{2010}, \emph{82},
  3045\relax
\mciteBstWouldAddEndPuncttrue
\mciteSetBstMidEndSepPunct{\mcitedefaultmidpunct}
{\mcitedefaultendpunct}{\mcitedefaultseppunct}\relax
\EndOfBibitem
\bibitem[Franke et~al.(2011)Franke, Schulze, and Pascual]{FSP2011}
Franke,~K.~J.; Schulze,~G.; Pascual,~J.~I. \emph{Science} \textbf{2011},
  \emph{940}, 332\relax
\mciteBstWouldAddEndPuncttrue
\mciteSetBstMidEndSepPunct{\mcitedefaultmidpunct}
{\mcitedefaultendpunct}{\mcitedefaultseppunct}\relax
\EndOfBibitem
\bibitem[Strozecka et~al.(2012)Strozecka, Soriano, Pascual, and
  Palacios]{SSP2012}
Strozecka,~A.; Soriano,~M.; Pascual,~J.~I.; Palacios,~J.~J. \emph{Phys. Rev.
  Lett.} \textbf{2012}, \emph{109}, 147202\relax
\mciteBstWouldAddEndPuncttrue
\mciteSetBstMidEndSepPunct{\mcitedefaultmidpunct}
{\mcitedefaultendpunct}{\mcitedefaultseppunct}\relax
\EndOfBibitem
\bibitem[Serri et~al.(2014)Serri, Wu, Fleet, Harrison, Hirjibehedin, Kay,
  Fisher, Aeppli, and Heutz]{SWF2014}
Serri,~M.; Wu,~W.; Fleet,~L.~R.; Harrison,~N.~M.; Hirjibehedin,~C.~F.; Kay,~C.;
  Fisher,~A.~J.; Aeppli,~G.; Heutz,~S. \emph{Nature Comm.} \textbf{2014},
  \emph{5}, 3079\relax
\mciteBstWouldAddEndPuncttrue
\mciteSetBstMidEndSepPunct{\mcitedefaultmidpunct}
{\mcitedefaultendpunct}{\mcitedefaultseppunct}\relax
\EndOfBibitem
\bibitem[Sessi et~al.(2013)Sessi, Otrokov, Bathon, Vergniory, Tsirkin, Kokh,
  Tereshchenko, Chulkov, and M.Bode]{SOB2013}
Sessi,~P.; Otrokov,~M.; Bathon,~T.; Vergniory,~M.; Tsirkin,~S.; Kokh,~K.;
  Tereshchenko,~O.; Chulkov,~E.; M.Bode, \emph{Phys. Rev. B} \textbf{2013},
  \emph{88}, 161407(R)\relax
\mciteBstWouldAddEndPuncttrue
\mciteSetBstMidEndSepPunct{\mcitedefaultmidpunct}
{\mcitedefaultendpunct}{\mcitedefaultseppunct}\relax
\EndOfBibitem
\bibitem[Uhlmann et~al.(2013)Uhlmann, Swart, and Repp]{USR2013}
Uhlmann,~C.; Swart,~I.; Repp,~J. \emph{Nano Lett.} \textbf{2013}, \emph{13},
  777\relax
\mciteBstWouldAddEndPuncttrue
\mciteSetBstMidEndSepPunct{\mcitedefaultmidpunct}
{\mcitedefaultendpunct}{\mcitedefaultseppunct}\relax
\EndOfBibitem
\bibitem[Ziroff et~al.(2010)Ziroff, Forster, Sch{\"o}ll, Puschnig, and
  Reinert]{ZFS2010}
Ziroff,~J.; Forster,~F.; Sch{\"o}ll,~A.; Puschnig,~P.; Reinert,~F. \emph{Phys.
  Rev. Lett.} \textbf{2010}, \emph{104}, 233004\relax
\mciteBstWouldAddEndPuncttrue
\mciteSetBstMidEndSepPunct{\mcitedefaultmidpunct}
{\mcitedefaultendpunct}{\mcitedefaultseppunct}\relax
\EndOfBibitem
\bibitem[Chang et~al.(2008)Chang, Kuck, Brede, Lichtenstein, Hoffmann, and
  Wiesendanger]{CKB2008}
Chang,~S.-H.; Kuck,~S.; Brede,~J.; Lichtenstein,~L.; Hoffmann,~G.;
  Wiesendanger,~R. \emph{Phys. Rev. B} \textbf{2008}, \emph{78}, 233409\relax
\mciteBstWouldAddEndPuncttrue
\mciteSetBstMidEndSepPunct{\mcitedefaultmidpunct}
{\mcitedefaultendpunct}{\mcitedefaultseppunct}\relax
\EndOfBibitem
\bibitem[Cuadrado et~al.(2010)Cuadrado, Cerd{\'a}, Wang, Xin, Berndt, and
  Tang]{CCW2010}
Cuadrado,~R.; Cerd{\'a},~J.; Wang,~Y.; Xin,~G.; Berndt,~R.; Tang,~H. \emph{J.
  Chem. Phys.} \textbf{2010}, \emph{133}, 154701\relax
\mciteBstWouldAddEndPuncttrue
\mciteSetBstMidEndSepPunct{\mcitedefaultmidpunct}
{\mcitedefaultendpunct}{\mcitedefaultseppunct}\relax
\EndOfBibitem
\bibitem[Baran and Larsson(2013)]{BL2013}
Baran,~J.~D.; Larsson,~J.~A. \emph{J. Phys. Chem. C} \textbf{2013}, \emph{117},
  23887\relax
\mciteBstWouldAddEndPuncttrue
\mciteSetBstMidEndSepPunct{\mcitedefaultmidpunct}
{\mcitedefaultendpunct}{\mcitedefaultseppunct}\relax
\EndOfBibitem
\bibitem[Zhang et~al.(2004)Zhang, Du, and Gao]{PSA2004}
Zhang,~Y.~Y.; Du,~S.~X.; Gao,~H.-J. \emph{Prog. Surf. Sci.} \textbf{2004},
  \emph{77}, 139\relax
\mciteBstWouldAddEndPuncttrue
\mciteSetBstMidEndSepPunct{\mcitedefaultmidpunct}
{\mcitedefaultendpunct}{\mcitedefaultseppunct}\relax
\EndOfBibitem
\bibitem[Cheng et~al.(2007)Cheng, Gao, Deng, Liu, Jiang, Lin, He, Du, and
  Gao]{CGD2007}
Cheng,~Z.~H.; Gao,~L.; Deng,~Z.~T.; Liu,~Q.; Jiang,~N.; Lin,~X.; He,~X.~B.;
  Du,~S.~X.; Gao,~H.-J. \emph{J. Phys. Chem. C} \textbf{2007}, \emph{111},
  2656\relax
\mciteBstWouldAddEndPuncttrue
\mciteSetBstMidEndSepPunct{\mcitedefaultmidpunct}
{\mcitedefaultendpunct}{\mcitedefaultseppunct}\relax
\EndOfBibitem
\bibitem[Mugarza et~al.(2010)Mugarza, Lorente, Ordej\'on, Krull, Stepanow,
  Bocquet, Fraxedas, Ceballos, and Gambardella]{MLO2010}
Mugarza,~A.; Lorente,~N.; Ordej\'on,~P.; Krull,~C.; Stepanow,~S.;
  Bocquet,~M.-L.; Fraxedas,~J.; Ceballos,~G.; Gambardella,~P. \emph{Phys. Rev.
  Lett.} \textbf{2010}, \emph{105}, 115702\relax
\mciteBstWouldAddEndPuncttrue
\mciteSetBstMidEndSepPunct{\mcitedefaultmidpunct}
{\mcitedefaultendpunct}{\mcitedefaultseppunct}\relax
\EndOfBibitem
\bibitem[Mugarza et~al.(2011)Mugarza, Krull, Robles, Stepanow, Ceballos, and
  Gambardella]{MKR2011}
Mugarza,~A.; Krull,~C.; Robles,~R.; Stepanow,~S.; Ceballos,~G.; Gambardella,~P.
  \emph{Nature Comm.} \textbf{2011}, \emph{2}, 490\relax
\mciteBstWouldAddEndPuncttrue
\mciteSetBstMidEndSepPunct{\mcitedefaultmidpunct}
{\mcitedefaultendpunct}{\mcitedefaultseppunct}\relax
\EndOfBibitem
\bibitem[Mugarza et~al.(2012)Mugarza, Robles, Krull, Koryt{\'{a}}r, Lorente,
  and Gambardella]{MRK2012}
Mugarza,~A.; Robles,~R.; Krull,~C.; Koryt{\'{a}}r,~R.; Lorente,~N.;
  Gambardella,~P. \emph{Phys.\ Rev.\ B} \textbf{2012}, \emph{85}, 155437\relax
\mciteBstWouldAddEndPuncttrue
\mciteSetBstMidEndSepPunct{\mcitedefaultmidpunct}
{\mcitedefaultendpunct}{\mcitedefaultseppunct}\relax
\EndOfBibitem
\bibitem[Zhang et~al.(2009)Zhang, Liu, Qi, Dai, Fang, and Zhang]{ZLQ2009}
Zhang,~H.; Liu,~C.-X.; Qi,~X.-L.; Dai,~X.; Fang,~Z.; Zhang,~S.-C. \emph{Nature
  Phys.} \textbf{2009}, \emph{5}, 438\relax
\mciteBstWouldAddEndPuncttrue
\mciteSetBstMidEndSepPunct{\mcitedefaultmidpunct}
{\mcitedefaultendpunct}{\mcitedefaultseppunct}\relax
\EndOfBibitem
\bibitem[Schlenk et~al.(2013)Schlenk, Bianchi, Koleini, Eich, Pietzsch,
  Wehling, Frauenheim, Balatsky, Mi, Iversen, Wiebe, Khajetoorians, Hofmann,
  and Wiesendanger]{SBK2013}
Schlenk,~T.; Bianchi,~M.; Koleini,~M.; Eich,~A.; Pietzsch,~O.; Wehling,~T.~O.;
  Frauenheim,~T.; Balatsky,~A.; Mi,~J.-L.; Iversen,~B.~B.; Wiebe,~J.;
  Khajetoorians,~A.~A.; Hofmann,~P.; Wiesendanger,~R. \emph{Phys. Rev. Lett.}
  \textbf{2013}, \emph{110}, 126804\relax
\mciteBstWouldAddEndPuncttrue
\mciteSetBstMidEndSepPunct{\mcitedefaultmidpunct}
{\mcitedefaultendpunct}{\mcitedefaultseppunct}\relax
\EndOfBibitem
\bibitem[Li et~al.(2010)Li, Li, Yang, and Hou]{LLY2010}
Li,~Z.; Li,~B.; Yang,~J.; Hou,~J.~G. \emph{Acc. Chem. Res.} \textbf{2010},
  \emph{43}, 954\relax
\mciteBstWouldAddEndPuncttrue
\mciteSetBstMidEndSepPunct{\mcitedefaultmidpunct}
{\mcitedefaultendpunct}{\mcitedefaultseppunct}\relax
\EndOfBibitem
\bibitem[Grobosch et~al.(2011)Grobosch, Mahns, Loose, Friedrich, Schmidt,
  Kortus, and Knupfer]{GML2011}
Grobosch,~M.; Mahns,~B.; Loose,~C.; Friedrich,~R.; Schmidt,~C.; Kortus,~J.;
  Knupfer,~M. \emph{Chem. Phys. Lett.} \textbf{2011}, \emph{505}, 122\relax
\mciteBstWouldAddEndPuncttrue
\mciteSetBstMidEndSepPunct{\mcitedefaultmidpunct}
{\mcitedefaultendpunct}{\mcitedefaultseppunct}\relax
\EndOfBibitem
\bibitem[Wang et~al.(2011)Wang, Zhu, Sun, Li, Zhang, Wen, Chen, He, Wang, Ma,
  Jia, Zhang, and Xue]{WZS2011}
Wang,~G.; Zhu,~X.-G.; Sun,~Y.-Y.; Li,~Y.-Y.; Zhang,~T.; Wen,~J.; Chen,~X.;
  He,~K.; Wang,~L.-L.; Ma,~X.-C.; Jia,~J.-F.; Zhang,~S.~B.; Xue,~Q.-K.
  \emph{Adv. Mat.} \textbf{2011}, \emph{23}, 2929\relax
\mciteBstWouldAddEndPuncttrue
\mciteSetBstMidEndSepPunct{\mcitedefaultmidpunct}
{\mcitedefaultendpunct}{\mcitedefaultseppunct}\relax
\EndOfBibitem
\bibitem[Koch et~al.(2014)Koch, Makarenko, Golyashov, Shegai, and
  Tereshchenko]{KMG2014}
Koch,~K.~A.; Makarenko,~S.~V.; Golyashov,~V.~A.; Shegai,~O.~A.;
  Tereshchenko,~O.~E. \emph{Cryst. Eng. Comm} \textbf{2014}, \emph{16},
  581\relax
\mciteBstWouldAddEndPuncttrue
\mciteSetBstMidEndSepPunct{\mcitedefaultmidpunct}
{\mcitedefaultendpunct}{\mcitedefaultseppunct}\relax
\EndOfBibitem
\end{mcitethebibliography}

\providecommand*{\mcitethebibliography}{\thebibliography}
\csname @ifundefined\endcsname{endmcitethebibliography}
{\let\endmcitethebibliography\endthebibliography}{}

\end{document}